\documentclass[%
superscriptaddress,
preprint,
 amsmath,amssymb,
 aps,
longbibliography]{revtex4-2}

\usepackage{graphicx}
\usepackage{dcolumn}
\usepackage{bm}
\usepackage{array}
\usepackage{cellspace}

\graphicspath{ {./} }
\usepackage{svg}
\usepackage{xcolor}
\usepackage{wasysym}
\usepackage{fdsymbol}
 \usepackage{multirow}
 \usepackage{tabularray}

\begin{document}

\preprint{
APS/123-QED}

\title{Receding contact line dynamics on superhydrophobic surfaces}

\author{Lorenzo Betti}
\affiliation{
Université Côte d’Azur, CNRS UMR 7010, Institut de Physique de Nice, 17 Rue Julien Lauprêtre, 06200 Nice, France}
\affiliation{Soft Matter Sciences and Engineering, PSL Research University, ESPCI Paris, Sorbonne Universite, CNRS UMR 7615, 10 rue Vauquelin, 75231 Paris Cedex 05, France}
\author{Jordy Queiros Campos}
\affiliation{
Université Côte d’Azur,  CNRS UMR 7010, Institut de Physique de Nice, 17 Rue Julien Lauprêtre, 06200 Nice, France}
\author{Amandine
 Lechantre}
\affiliation{
Université Côte d’Azur,  CNRS UMR 7010, Institut de Physique de Nice, 17 Rue Julien Lauprêtre, 06200 Nice, France}
\author{Léa Cailly-Brandstater}
\affiliation{
Université Côte d’Azur,  CNRS UMR 7010, Institut de Physique de Nice, 17 Rue Julien Lauprêtre, 06200 Nice, France}
\author{Sarra Nouma}
\author{Jérôme Fresnais}
\affiliation{Physicochimie des Électrolytes et Nanosystèmes Interfaciaux, Sorbonne Université, CNRS, F-75005 Paris, France}
\author{Etienne Barthel}
\affiliation{Soft Matter Sciences and Engineering, PSL Research University, ESPCI Paris, Sorbonne Universite, CNRS UMR 7615, 10 rue Vauquelin, 75231 Paris Cedex 05, France}
\author{Yann Bouret}
\author{Xavier Noblin}
\affiliation{
Université Côte d’Azur,  CNRS UMR 7010, Institut de Physique de Nice, 17 Rue Julien Lauprêtre, 06200 Nice, France}
\author{Céline Cohen}
\email{celine.cohen@univ-cotedazur.fr}
\affiliation{
Université Côte d’Azur,  CNRS UMR 7010, Institut de Physique de Nice, 17 Rue Julien Lauprêtre, 06200 Nice, France}

\date{\today}

\begin{abstract}
    { 
We have explored receding contact line dynamics on superhydrophobic surfaces, composed of micropillars arrays. We present here dynamic receding contact angle measurements of water on such surfaces, covering contact line speeds spanning over five decades. We have studied the effect of pillars fraction on dynamical receding contact angles. We compared these measurements to those on smooth surfaces with the same chemical nature and also with similar systems reported in the literature.
 
We show that superhydrophobic surfaces exhibit a significantly lower dependence of contact angle on contact line speed compared to smooth surfaces. Additionally, we observed that a higher surface fraction of pillars leads to a greater dependence of the contact angle on contact line speed, approaching the dependence of the angle on smooth surface. Interestingly, we show that the exact texuration of the surface does not play a fundamental role in the angle-velocity relationships as long as microtextures present the same type of periodic pattern (pillar arrays or microgrid). These results are interpreted in terms of viscous friction reduction on superhydrophobic surfaces, shedding light on the underlying mechanisms governing their unique dynamic behavior.
In addition we show that contact angles follow same laws for two different geometries (milimetric sessile drop and a centimetric capillary bridge). 
}
\end{abstract}

\maketitle

\section{\label{sec:level1} Introduction}
Wetting phenomena are used in most industrial areas (coatings, food, aviation, etc.). They have been the focus of intensive academic research for decades especially superhydrophobic surfaces that present remarkable anti adhesion properties \cite{roach2008progess, nosonovsky2009superhydrophobic}.
They are mainly characterized by measuring the equilibrium contact angle ($\theta_e$), which quantifies the ability of the liquid to spread on a solid. From a purely theoretical point of view, this angle results from a balance of the surface tensions of the three interfaces involved (liquid-solid, solid-gas, liquid-gas) \cite{de1985wetting}. In practice, however, it has been established that this angle depends on the direction of movement of the liquid on the substrate (receding or advancing), as the roughness of the surfaces can trap the contact line. These pinning lead to a deformation of the contact line locally on the defects, but also to a deformation of the liquid interface (liquid-gas) in the plane perpendicular to the solid surface, which is also deformed over a characteristic distance comparable to the distance between the defects (fringe elasticity) \cite{de1985wetting}. These deformations, which vary according to the direction of movement of the contact line, lead to different contact angles for a line moving forward on a substrate ($\theta_{a}$) and for a receding line ($\theta_{r}$). This phenomenon is known as contact angle hysteresis \cite{butt2022contact}. To be precise, the contact angle hysteresis is the difference between the quasi-static limit of the angles: $\theta_{a,s
}$ and $\theta_{r,s}$. Research on the hysteresis of superhydrophobic surfaces has advanced significantly over the past decades, with a universal description recently proposed by Rivetti \textit{et al.} \cite{rivetti2015surface}. Studies show that for strong roughness, the contact line moves in a jerky manner due to surface defects that constantly trap and release it. McHale \textit{et al.} observed that these trapping events cause oscillations in the contact angle \cite{mchale2005analysis}, while Gauthier \textit{et al.} demonstrated that the shape of micropillar arrays influences the movement of the triple line \cite{gauthier2013role}. In 2015, Rivetti \textit{et al.} modeled the "zipper" movement of contact lines on micropillar arrays, showing a linear increase in the receding contact angle with surface fraction \cite{rivetti2015surface}.
We have adapted recently the capillary bridge setup \cite{restagno2009contact} to the study of superhydrophobic surfaces \cite{cohen2019capillary}. With, this setup, in the quasi-static regime, it is possible to measure receding and advancing contact angles with an indirect measurement that enables to get the angles from the measurement of contact area instead of the direct contact angle measurement on profile images. This allows to overcome the difficulties of direct measurement, which leads to underestimation of contact angles for high contact angles presented by superhydrophobic surfaces, as shown by Schellenberger \textit{et al.} \cite{schellenberger2016water}. We have then been able to measure higher contact angles which permit us to show that the dependence of receding contact angle with pillar fraction is well explained with local deformation models \cite{dubov_elastic_2012, rivetti2015surface} but it is not the case for advancing contact angle.   
In these different works on superhydrophobic surfaces, the contact line is assumed to be at the limit of motion, at very low speed and the effects of the speed of motion on these deformations have not been studied yet. In addition, even though the variation of the contact angle $\theta_d$ with the line speed  $\theta_d(V_l)$ has been extensively studied in the literature for smooth surfaces \cite{mohammad2022review, ramiasa2011contact, rio_boundary_2005, rolley_dynamics_2007, hoffman1975study, snoeijer_moving_2013}, the laws for superhydrophobic surfaces are still poorly understood and little researched. For smooth surfaces and viscous liquids, experiments show a linear dependence of the cubic dynamic contact angle on the velocity \cite{rio_boundary_2005,hoffman1975study}. This dependence was modeled by Cox-Voinov using a hydrodynamic approach, which assumes that the capillary energy is dissipated by the viscosity \cite{voinov_hydrodynamics_1976, cox_dynamics_1986}. In these models, it is assumed that the contact angles are small compared to 1. Other experiments have been performed with low-viscosity and highly wetting liquids such as liquid helium \cite{rolley_dynamics_2007}, which in this case indicates contact line dynamics driven by thermally activated molecular adsorption and desorption processes \cite{blake1969kinetics}. This second approach is known under the name of MKT model (Molecular Kinetic Theory) \cite{blake1969kinetics}. 
More recently, models combining these two approaches have also been proposed, either by considering the molecular processes using modified characteristic lengths in the hydrodynamic model \cite{limat2020dewetting} or by introducing an effective viscosity in the MKT approach to capture the hydrodynamic dissipation in this model \cite{duvivier2011experimental}. In \cite{limat2020dewetting}, authors have shown for a sliding droplet that viscous dissipation at the corner that forms at the back of the droplet can have a particularly strong influence on the numerical fitting parameters.

Despite all these studies, understanding and modeling the dynamics of the contact angle still remains an open problem today. Indeed, hydrodynamic modeling seems to explain well certain measurements of contact angles for wetting surfaces and viscous liquids \cite{rio_boundary_2005}, but does not allow to explain the data in a universal way, in particular for complex contact geometries such as those of plates pulled out of a liquid bath \cite{snoeijer_moving_2013}. The dynamics of superhydrophobic surfaces are also still poorly understood. There are only a few contact angle measurements on these complex surfaces in the literature and, to our knowledge, no measurements with pure water \cite{karim_experimental_2018, kim_dynamic_2015, harikrishnan2018correlating}. The first studies show dependencies of contact angles on the speed of the contact line that differ from those observed on smooth surfaces. Furthermore, they do not show universal behaviors. The obtained laws $\theta_d \ (V_l)$ depend on the texturing of the surfaces, the size and shape of the textures \cite{karim_experimental_2018} and the nature of the liquids. The work of Ramiasa \textit{et al} \cite{ramiasa2011contact}, Kim \textit{et al} \cite{kim_dynamic_2015} and Mouterde \textit{et al} \cite{mouterde2019superhydrophobic} suggest that the presence of a slippage on the surface of these textured substrates could be one of the key parameters to understand contact angles dynamics, but no model is currently proposed.
Finally, the results presented in the literature are difficult to interpret because the experimental parameters are often coupled and the liquids used are complex fluids, as can be seen in Table \ref{tab1}, which summarizes the main studies on superhydrophobic surfaces.
 \begin{table}
    \begin{center}
        \begin{tblr}{
          colspec={XXXX},
          cells={valign=m,halign=c},
          vlines,hlines
        }
        Articles & Ca & System & Wetting state \\
        Karim \textit{et al} (2018) \cite{karim_experimental_2018} (experimental) & $3.10^{-4} - 4.5.10^{-3}$ & Teflon/PEG & Wenzel\\
        Kim \textit{et al} (2015) \cite{kim_dynamic_2015} (experimental)  & $1.10^{-3} - 2.10^{-1}$ & Viscous solution / superhydrophobic surfaces  & Cassie \\
        Harikrishnan \textit{et al} (2018) \cite{harikrishnan2018correlating} (experimental) & $5.10^{-7} - 5.10^{-6}$ & Complexe fluids / superhydrophobic surfaces & - \\
        Lee \textit{et al} (2022) \cite{lee_contact_2022} (numerical) & - & - & Cassie \\
        \end{tblr}
    \end{center}
\caption{Dynamic contact angle measurement in the literature .}\label{tab1}
    \end{table}

To better understand these different behaviors, we conducted model experiments in which we systematically varied only one parameter at a time. We investigated the relationship between the contact angle and the speed of the contact line for a receding line on surfaces covered with micropillars whose size and surface fraction were varied. To do so, we performed measurements with two different devices (at millimetric and centimetric scales) to study the effects of contact geometry on this relationship. First, at millimetric scale using a sessile droplet setup mounted on an inverted microscope and second, at centimetric scale using the "capillary bridge method", which allows to explore larger wetted contact areas (closer to applications) and larger contact line speeds than with the sessile droplet setup \cite{restagno2009contact, cohen2019capillary}. In both setup, we force the displacement of the contact line on the surface at a chosen speed and measure the resulting receding contact angle and the contact line velocity. We compared the measurements on superhydrophobic surfaces with those on smooth surfaces of the same chemical nature and with results from the literature on similar systems.

We first present the experimental setups, the surface properties and the image analysis method. Then we present our results on the influence of the surface fraction on the dynamic receding contact angle for water on superhydrophobic surfaces for each method. Finally, we confront these results with the hydrodynamic and molecular kinetic approach as well as with a phenomenological friction law.

\section{Materials and Methods}

    \begin{figure}[h]
    \includegraphics[scale = 1]{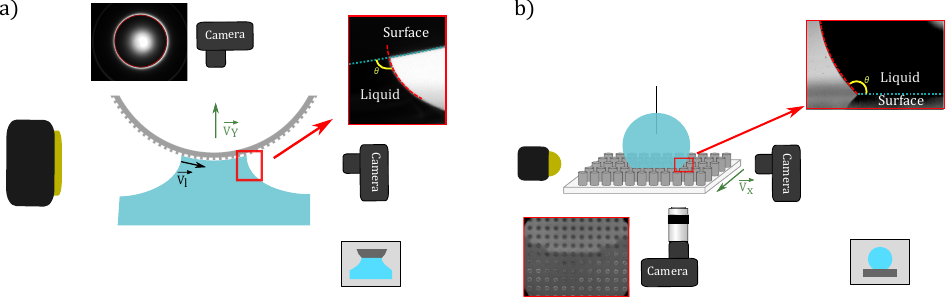}
    \caption{Schematic view of a) the capillary bridge setup (setup 1) and b) the sessile droplet setup (setup 2).}
    \label{montage}
    \end{figure} 
    The figure \ref{montage} presents the two different setups used to measure the receding contact angles for a given contact line speed : (a) the capillary bridge setup that is inspired by the setup used by Cohen {\it{et al.} }\cite{cohen2019capillary} and (b) a sessile drop setup mounted on an inverted microscope. For both setups, contact line is forced to recede on the solid at a given velocity with a motorized micrometric stage. 
    The capillary bridge method consists in following the deformations of a capillary bridge formed between the solid surface (on top) and a liquid bath when the solid is displaced vertically.
    The vertical translation of the surface is ensured by a worm screw connected to a stepper motor controlled by a LabVIEW program developed in the laboratory. The device allows to control the vertical movements of the stage over a distance of 10 cm with an accuracy of a few micrometers for translation speeds between 1 and 5000$\mu m/s$. First, the solid surface is approached vertically from the free surface of the liquid bath until contact is made and a capillary bridge is created, at a speed of 1$\mu m/s$. Then the surface is pushed in the liquid bath in order to increase the contact area at the same speed. After that, it is pulled up at a constant speed $V_{y}$ until the liquid bridge breaks.
    During these steps, we record the capillary bridge profile with a high speed camera (Phantom V7.11) positioned horizontally and we measure the contact angle and the contact line displacements. The film from the high-speed camera was obtained between 30 and 5000 fps, depending on the surface translation speed.
    
    In case of the sessile drop setup, an horizontal high-speed camera (Phantom V7.3)) is used to measure the advancing and the receding contact angles of a drop helded by a syringe tip while the surface is moved horizontally by a high-speed micrometric stage connected to a stepper motor (Fig. \ref{montage} a). The horizontal movements are controlled over a distance of 50 mm with an accuracy of a few micrometers at translation speeds of up to 50 mm/s. A second camera is connected to the inverted optical microscope to observe the contact surface between the drop and the solid surface.

\begin{figure}[h!]
    \includegraphics[scale = 1]{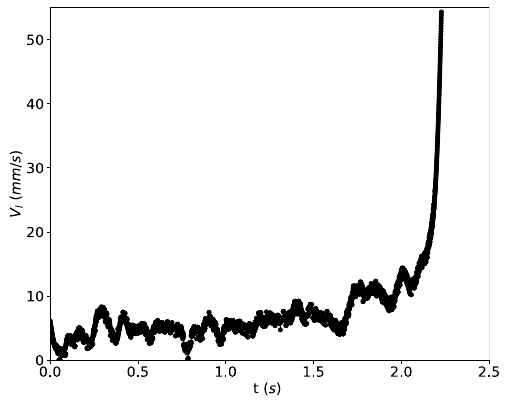}
    \caption{Speed of the contact line as a function of time during the pulling process on the capillary bridge.}
    \label{Vligne}
    \end{figure}

    The image analysis of both experimental setups is carried out with a Python script. From the images, we subtract a fixed value from all the gray levels in order to remove the various reflections present in the liquid. Then, we apply a Canny filter to detect the contour that separates the air from our system. The pixels corresponding to the liquid (red) and the solid substrate (blue) are identified on the Fig.\ref{montage}. The contact between the liquid and the substrate is determined by the intersection of the two parts and represents the position of the contact line. The contact angle $\theta$ is estimated by calculating the angle formed between the tangents of each circles at the point of intersection illustrated on Fig.\ref{montage}. As mentionned in the introduction, the main difficulty and source of error for direct measurement of contact angles lies in the determination of the triple point. Because of the resolution of the image and the presence of shadows, the exact position of the triple point cannot be known with a great accuracy \cite{schellenberger2016water}. The position of the triple point is therefore estimated at plus or minus 4 pixels. This script allow us to measure the contact angle with an uncertainty of $\pm 3.5^{\circ}$.
    \\ 
    In the capillary bridge setup, the instantaneous contact line speed  $V_l$ is obtained by deriving the position over time. It is important to note that a constant velocity of pulling, $V_{y}$, does not correspond to a constant contact line speed $V_{l}$. First the speed of the line increases almost linearly, and then it accelerates rapidly. This is due to the curved geometry of the substrate and the breakage of the capillary bridge combined with the pulling speed. By fixing the pulling velocity $V_{y}$, we have access to a large range of contact line speed $V_{l}$ as shown in Fig.\ref{Vligne}. In contrast, in the sessile drop setup, the contact line speed is fixed and it is equal and opposite to the surface translation speed $V_{x}$.
    \\
    The surfaces studied for the capillary bridge experiments are curved with a curvature radius $r_c = 81 \, mm$ and transparent. All the superhydrophobic surfaces (SH) used in this study are made with the same protocol of photo-lithography used by Cohen \textit{et al }\cite{cohen2019capillary}. They are made of a square array of cylindrical micro metric pillars in Norland Optical Adhesive (NOA 81) coated with PTFE thin films (spin coated Teflon AF 1601). The surfaces used for sessile drop setup are similar but the textures and coatings are made on flat glass slides. The geometric properties of all surfaces are sum up in the Table \ref{tab2}.

    \begin{table}
    \centering
    \begin{tabular}{ |c|c|c|c|c|c| }
        \hline
         & Surface & $d$ ($\mu m$) & $e$ ($\mu m$) & $h$ ($\mu m$) & $\phi$\\
        \hline
        \multirow{4}{*}{Curved surfaces} & SH  & 29.9 & 70 & 100 & 0.14\\
        & SH & 11.9 & 25 & 25 & 0.18\\
        & SH & 10.8 & 20 & 25 & 0.23\\
        & Smooth & - & - & - & 1\\
        \hline

        \multirow{3}{*}{Flat surfaces} & SH & 27.0 & 70 & 100 & 0.12\\
        & SH & 10.4 & 25 & 25 & 0.14\\
        & SH & 10.0 & 20 & 25 & 0.20\\
        
        \hline
    \end{tabular}
    \caption{Surfaces geometric properties. For the superhydrophobic surfaces (SH), $d$ is the diameter of the pillars, $e$ is the space between two pillars, $h$ is the height of the pillars and $\phi$ is the solid fraction of the pillars $\phi=\frac{\pi d^2}{4e^2}$.}\label{tab2}
    \end{table}
\section{Results and discussion}

    \begin{figure}[h!]
    \includegraphics{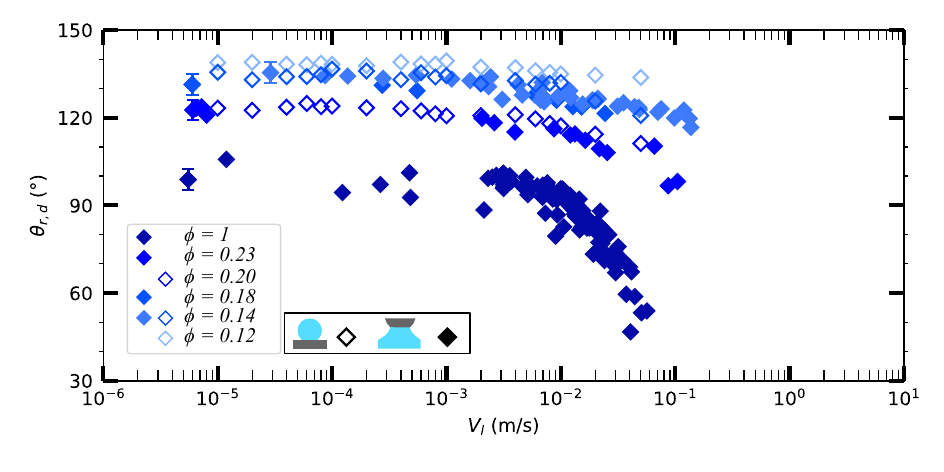}
    \caption{Receding contact angle as function of the contact line speed. The closed and open symbols represent respectively measurements obtained with capillary bridge and sessile drop setups. The blue gradient represents the data for different solid fractions from the smooth surface (darkest) to the smaller $\phi$ (lightest)}
    \label{theta_V_all}
    \end{figure}

    Figure \ref{theta_V_all}  illustrates the measurements of the receding contact angles as a function of the contact line speed obtained with the two experimental setups for water on PTFE coated surfaces. The results are represented by a blue gradient from the darkest for the smooth surface to increasingly lighter shades for superhydrophobic surfaces made up of micropillars spaced further and further apart (corresponding to a decreasing solid surface fraction). 
    
    Firstly, we can notice that for all the substrates, at low contact line speeds, the contact angle remains nearly constant, corresponding to the static receding angle $\theta_{r,s}$. Then, the contact angle decreases with the contact line speed above a certain threshold in velocity ($\approx 2$ mm/s). This threshold does not depend on the surface coating. 
    
    Secondly, we can see that the smooth surface exhibits a rapid decrease in the contact angle as the speed increases, compared to the superhydrophobic surfaces where the contact angle shows a weak dependence on the contact line speed. The graph also indicates that the variation in the contact angle with speed becomes progressively less significant as the solid fraction decreases. Interestingly, the measurements obtained with both experimental setup are very similar and thus highly complementary, reinforcing the reliability and validity of the observed trends.
    
\begin{figure}[h!]
    \includegraphics[scale = 1]{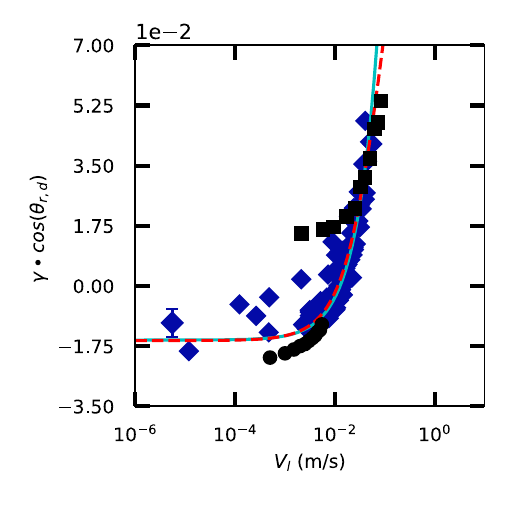}
    \caption{Receding capillary force (per unit length) as function of the contact line speed of water on a smooth PTFE surface measured with setup 1 (blue). The black symbols correspond to results from  for comparable systems in the literature (circle \cite{wang_determination_2020} and square \cite{kim_dynamic_2015}). The full and dash lines represent the fits of  the hydrodynamical and molecular kinetics theory models.}
    \label{theta_V_smooth}
    \end{figure}

\paragraph{Analysis of smooth surfaces.}
Figure \ref{theta_V_smooth} presents results of smooth surfaces plotted this time in terms of the receding capillary force per unit length 
$\gamma \cos\theta_{r}$ as function of contact line speeds. 

The black points correspond to results from literature on similar systems i.e. water on PTFE coated smooth surfaces but with different measurement methods \cite{kim_dynamic_2015, wang_determination_2020}.   
The quasi-static receding contact angle ($\theta_{r,s}$, mean of the points of the plateau obtained at low velocities) presents relative dispersed values. This dispersion can be explained both by the high sensitivity of the receding contact angle on surfaces defects which can be very different for all surfaces presented in this graph and on the nature of the PTFE used for the different coating. However, the results are very similar for all studies for the dynamical regime. We have used the classical models from the literature to analyse these data. Regarding the scale of our systems which are macroscopic (milimetric for the drop and centimetric for the capillary bridge), the hydrodynamic model is the more adapted approach to describe our experiments. It predicts the following relation for the contact angle : 
\begin{equation}
    \gamma\cos\theta_{r,d}= \gamma\cos\theta_{r,s} - 3\eta V_l \ln \left(\frac{L}{a}\right)
\label{eq1}
\end{equation}
where, $\gamma$ is the surface tension, $\eta$ is the viscosity and $a$ and $L$ are respectively the smaller and the higher characteristic distance of the system. $\theta_{r,s}$ is estimated from the mean of the values of the quasi-static regime plateau and $a$ and $L$ are left as free parameters. The full line represents the fit obtained for $\theta_{r,s}= 103 ^{\circ}, \ln \left(\frac{L}{a}\right) = 8.3.10^4$. We can notice that the model fits well the data. However, the adjustable parameters used have no physical viability. Indeed, the $\ln$ argument needs to be around $10^8$ which imposes an unreasonable value for the molecular size $a$. 

It has already been shown in literature that the hydrodynamic model does not adequately describe the dynamics of the contact angle in the case of dip coating \cite{snoeijer_moving_2013}. We show that it is also not applicable quantitatively for the geometry of the capillary bridge. 
 
We have then applied the MKT model which predicts the following dependency for the contact angle versus contact line speed \cite{blake1969kinetics} : 
\begin{equation}
    \gamma\cos\theta_{r,d}= \gamma\cos\theta_{r,s}+\frac{2 k_B T}{\lambda^2} \sinh^{-1}\frac{V_l}{2K_w\lambda}
\label{eq2}
\end{equation}
$\theta_{r,s}$ is estimated as explained above and $K_w$ and $\lambda$ are left as free parameters. They respectively represents the frequency of molecular jumps and the distance between to adsorption/desorption site. $k_B$ being the Boltzman constant. The dash line represents the fit obtained with $\theta_{r,s}=103, K_w=4.8.10^7 \, Hz, \lambda= 0.39 \, nm $. 
We can see that this model presents a good agreement with the experimental data with fitting parameters comparable to the values reported in literature \cite{ramiasa2011contact}. However, it is important to notice that due to experimental uncertainties, there is a high variability on fitting parameters that allow to fit the data as reported in supplemental materials (section \ref{sup mat}). In addition, to be fully rigorous, this model should be used only for contact line at smaller scale because it supposes that the contact line movements are due to thermally activated molecular jumps which have no reason to be the case of our macroscopic experiments. Indeed, by calculating the thermal length $l_T=\sqrt{\frac{k_BT}{\gamma}}$ which compares thermal effects and capillary effects, we find that to have the molecular effects higher than the capillary ones, the systems should have a typical size below $0,24$ nm (expected range of size for validity of the MKT model).
Finnaly, for water on smooth surfaces, both models allows to fit the data but the results do not permit to well clarify the physical behaviors involved. However, as it is explained in the next section, those approaches are equivalent to a phenomenological friction model that can be used to interpret those results.  

\begin{figure} [h!]
    \includegraphics[scale = 1]{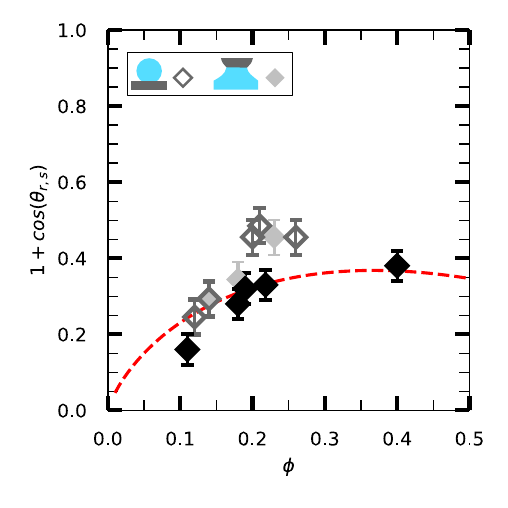}
    \caption{Effective adhesion for the quasistatic regime as a function of surface fraction $\phi$.}
    \label{theta_phi}
    \end{figure}

\paragraph{Analysis of superhydrophobics surfaces and comparison with smooth surface.} First, as we can see on figure \ref{theta_V_all}, the static receding contact angle increases as the solid fraction decreases. This implies an increase of the effective adhesion  $1+\cos\theta_{r,s}$ as function of the pillar fraction as it is shown in figure \ref{theta_phi}. Theses results are consistent with previous work of our team with the same experimental system for the quasistatic regime \cite{cohen2019capillary}. The dash line represents the analytical model of Jansons \cite{jansons1985moving} that has shown that the contact line elasticity model enables to explain the receding contact angle dependency with $\phi$ in term of liquid interface deformations due to pinning of the contact line on pillars while the contact line recedes \cite{rivetti2015surface, jansons1985moving, cohen2019capillary}.

\begin{figure} [h!]
    \includegraphics[scale = 1]{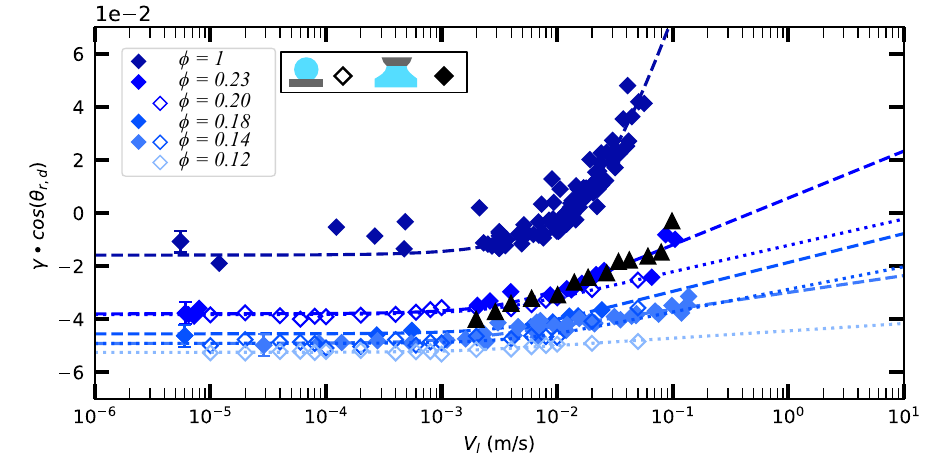}
    \caption{Receding contact angle as function of the contact line speed. The closed and open symbols represent respectively measurements obtained with capillary bridge and sessile drop setups. The blue gradient represents the data for different solid fractions from the smooth surface (darkest) to the smaller $\phi$ (lightest). The dash lines are the MKT fits of each curve with fitting parameters reported in table \ref{tab3}.}
    \label{theta_V_SH}
    \end{figure}

Secondly, for the dynamical regime, as observed for smooth PTFE surfaces, the hydrodynamic model fits also the data of superhydrophobic surfaces with fitting parameters presenting non physical order of magnitude. We have then focus the analysis of these date with the MKT model for wich we plot the fits in Figure \ref{theta_V_SH} (a). 
Table.\ref{tab3} shows the values of the fitting parameters used  for each surface. 
The fitting parameters are plotted as function of pillar fraction in figure \ref{theta_V_SH} (b).
 
      \begin{figure} [h!]
    \includegraphics[scale = 1]{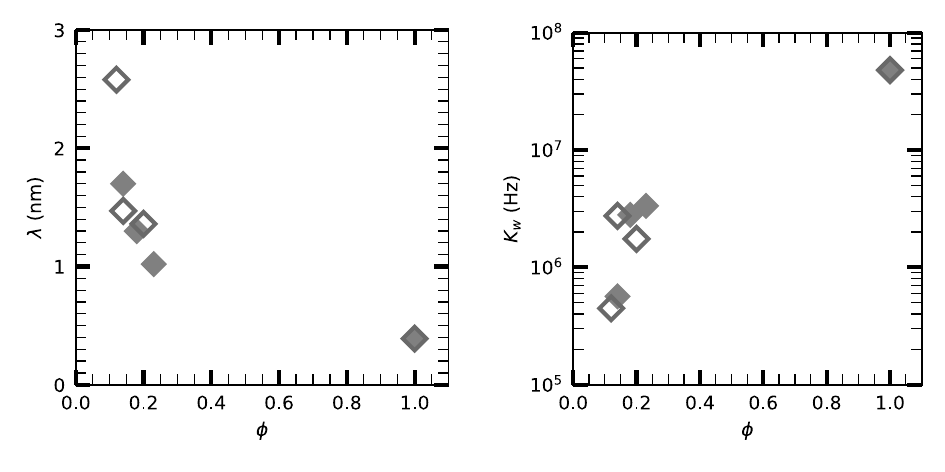}
    \caption{Fitting parameters of MKT fits : (left) the distance between adsorption sites $\lambda$ and (right) the molecular jumps frequency, as function of the solid fraction $\phi$.}
    \label{val_mkt}
    \end{figure}
    
 For the smooth surface, the average molecular displacement ($\lambda$) gives the lowest value and the frequency of molecular jumps ($K_w$)  is the highest compare to the superhydrophobic surfaces. For the microtexturated surface we observe that $\lambda$ increase and as the same time $K_w$ decrease when the solid fraction decrease. In addition, fitting parameters evolves monotonously with $\Phi$. This result can be interpreted in a phenomenological point of view. Indeed, if we considered that $\lambda$ is linked to the macroscopic jump from a pillar to another, it should increase if the distance between pillar increases so when pillar fraction decreases. For the same reason, the larger the solid fraction, the higher the jump frequency.
In addition, it is interesting to notice that the linearization of MKT model leads to a linear dependency of the capillary force with the contact line speed which is equivalent to a simple phenomenological friction model \cite{ren2007boundary}. This approach predicts the following dependency for the contact angle : 

\begin{equation}
     \gamma\cos\theta_{r,d}= \gamma\cos\theta_{r,s}+\beta V_l
     \label{eq3}
\end{equation}
where $\beta$ is the effective friction coefficient which is left as a free parameter for fitting the data. 

\begin{figure}
    \includegraphics[scale = 0.95]{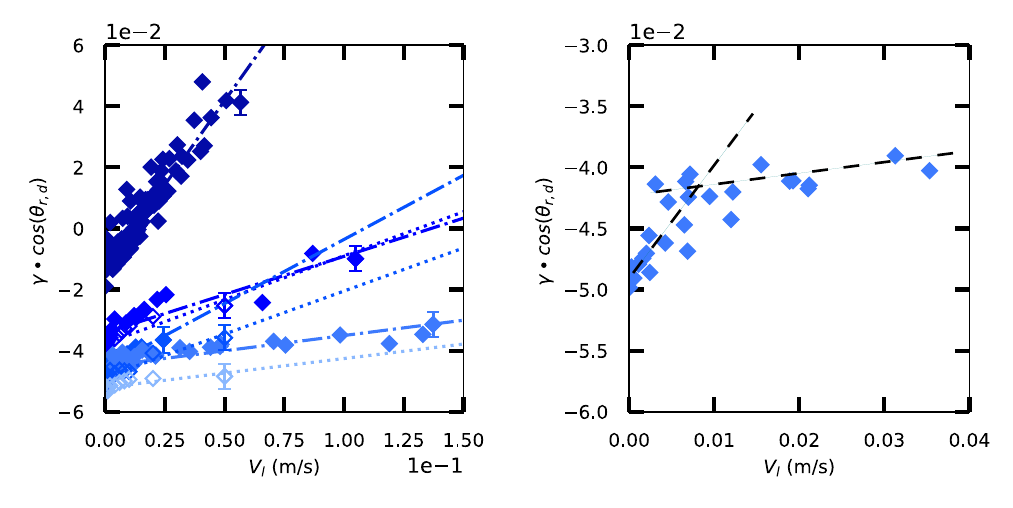}
    \caption{Receding capilarry force (per unit of length) as function of the contact line speed. The closed and open symbols represent respectively measurements obtained with capillary bridge and sessile drop setups. The blue gradient represents the data for different solid fractions from the smooth surface (darkest) to the smaller $\phi$ (lightest). The dash lines are the linear fits of each curve. On the right : a zoom for the smaller contact line speed for a typical curve. black thin dash lines are guide for eyes.}
    \label{REN}
    \end{figure}

We have also used this approach to fit the data. We have then plotted the capillary force as function of contact line speed in a linear-linear plot in figure \ref{REN} (a).

We can see that the data can be fitted by lines of equation \ref{eq3} for all surfaces. 
In addition, we can see on a zoom of a particular surfaces \ref{REN} (b), that the slope is higher for the lower speeds. We have neglected this difference when fitting the data with equation \ref{eq3} that has been applied to all the points. Indeed, doing the fit on only the last points changes the obtained coefficients from $7\%$ up to $20\%$ and it does not change the trend of the dependency of $\beta$ with $\Phi$.

\begin{figure}
    \includegraphics[scale = 1]{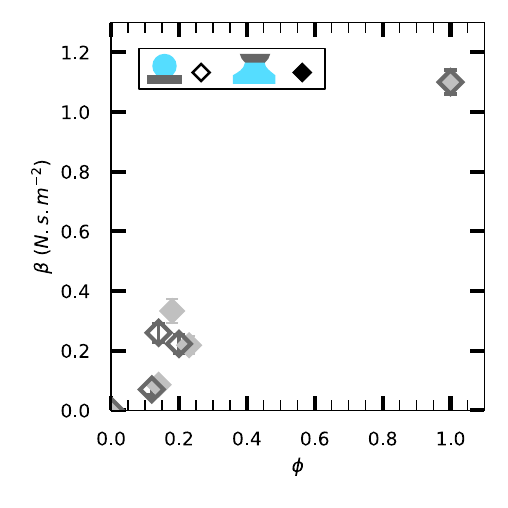}
    \caption{Effective friction coefficient as function of solid fraction.}
    \label{Beta}
    \end{figure}

Figure \ref{Beta} presents the effective coefficient $\beta$ obtained by linear fits for all data as function of the pillar fraction. It shows that the effective friction coefficient decreases when the solid fraction decreases. This can be interpreted as a reduction of friction due to the reduction of solid area of contact for superhydrophobic surfaces. 

 Moreover, we observe that the weak dependence of the contact angles on the contact line velocity is persistent up to velocities of several meters per second. Even at these high speeds, we do not observe a dynamic transition to a liquid entrainment regime as observed for smooth surfaces in this speed range. Indeed, when we observe the last moments of life of the capillary bridge, where the speed is the greatest, we do not observe any pinch-off of macroscopic water drops on the superhydrophobic surfaces unlike the smooth surfaces studied. This result is also in line with the work of Mouterde et al \cite{mouterde2019superhydrophobic} who observed that the shape of a drop rolling down an inclined microtextured superhydrophobic surface, at several meters per second, remained very close to its quasistatic shape. This also suggests that viscous friction is greatly reduced on these surfaces.

\begin{figure}
    \includegraphics[scale = 1]{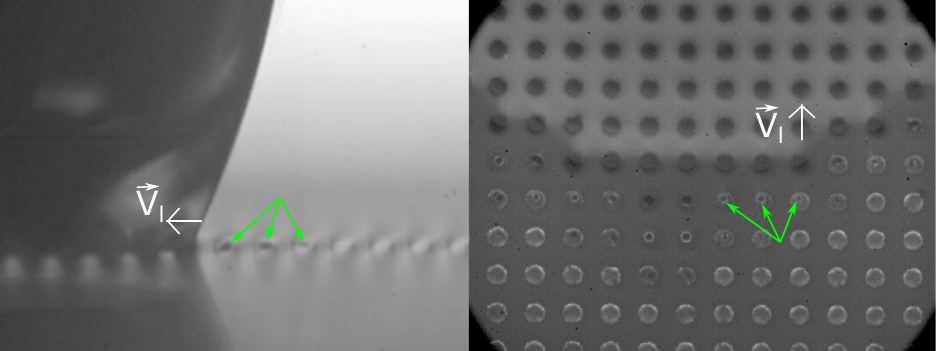}
    \caption{Images of a drop from side (a) and bottom (b) views while the contact line has receded on the surfaces. Green arrows shows droplets left on pillars during the contact line receding.}
    \label{goutte}
    \end{figure}

In addition, the two values of $\beta$ obtained for $\phi=0.14$ have been obtained for surfaces with the same pillar fraction but for two different pillar diameters (10 µm and 28 µm). We can see that there is a factor three between friction coefficients interestingly for a factor three between this two diameters. This effective coefficient friction is higher when the diameter is smaller so when there is more pillars. It suggests that even if friction is reduced on pillars surfaces, it is higher for small pillars that suggest that there is another effect to take into account to explain the higher value of $\beta$ for smaller pillars. We think that pillars are defects on which contact line can be pinned and deformed and eventually causes the pinch off of small droplets when contact line jump from a pillar to another as it can be seen on side and bottom view images in figure \ref{goutte}. Indeed, while the contact line is trapped on a micropillar, it forms a small capillary bridge. When the line moves, the liquid bridge breaks off and leaves a small droplet attached to the top of the pillar. 

This pinch off is also responsible of energy dissipation and for a given solid fraction, this effect can be higher for smaller pillars because there is more pinning sites. We think then that to explained theoretically the data of figure \ref{theta_V_SH}, it is needed to take into account, the dissipation due to friction on solid parts, the elastic deformation due to the pinning of contact line on pillars and the pinch off of droplets. 

 We have also compared our results with a recent study \cite{kim_dynamic_2015} carried out with a mixture of water and glycerol on a textured PTFE surface (micrometric grooves in the form of a grid). These data are plotted in black on graph \ref{theta_V_SH}. Their measurements show a dependency close to that obtained for our surfaces. This suggests that the influence of surface microtextures arrangement on the dependence of dynamic contact angle on velocity, remains of limited importance on the overall trend. In particular, the three textures studied have one important common point : the micro-metric textures array present a "square" arrangement of the textures. This is coherent with results obtained by  Gauthier \textit{et al.} \cite{gauthier2013role} in quasi-static conditions. Indeed, they have shown that the contact line retraction mechanism on this type of lattice will follow a similar "zip" closure mechanism contrary to axisymmetric array that presents another retractation mecanism. These different mecanisms have been explained by the local deformation of contact line on microtextures during the receding which are the sames for periodic arrays what ever the precise array parameters.  This could therefore explain why, for the three different texturations with similar arrays compared in figure \ref{theta_V_SH}, we obtain a close trend for the dynamic contact angles. This confirm the importance to consider the local contact line deformations on surface defects in the modelisation of the $\theta_d (V_l)$ law for superhydrophobic surfaces.

\section{Conclusion}
In this study, we explored for the first time with a model system (water on controlled microtextured surfaces), the relationship between contact angle and velocity of a receding contact line on hydrophobic and superhydrophobic surfaces over five decades on speeds. With the chemical nature kept identical between smooth and microtextured surfaces, we were able to study the effect of texturing on the $\theta_{r,d}-V_l$ law. More precisely, we used surfaces covered with micropillars arranged in a square array, and we varied the surface fraction of micropillars. We were thus able to show that the variation of the receding contact angle with contact line speed is much smoother for superhydrophobic surfaces than for non textured surfaces of the same chemical nature. Moreover, this dependence is stronger for larger surface fraction, and it increases continuously as we approach the non textured case. We have interpreted this result in terms of a reduction in liquid friction due to the reduction in real contact area on textured surfaces. In addition, our experiments highlight the role  on the $\theta_{r,d}-V_l$  law of pinning and deformations of the contact line, as well as the possible pinch off of micro-droplets on pillars. Finally, we have shown that the classical models  (hydrodynamic approach, molecular approach) need to be modified to take into account the changes in effective liquid friction. Indeed, on superhydrophobic surfaces, direct liquid friction should occur only on the top of the pillars. In addition, at this direct liquid friction contribution, the local deformations of contact line pinned on pillars as well as the energy dissipated by the pinch off of droplets add another contribution to the total effective liquid friction that should also taken into account.


\section{Acknowledgments}

We would like to thank Yaroslava Izmaylov for her support in the fabrication of superhydrophobic surfaces. 
We thank the ANR forfunding these researches within the ANR Project MADNESS (ANR-19-CE06-0023).

\bibliography{biblio_impact}
\newpage
\section{Supplemental Material}

\label{sup mat}

    \begin{table}[h!]
    \centering
    \begin{tabular}{ |c|c|c|c|c| }
        \hline
        & Surface & $\theta_{s,r}$ & $\lambda$ ($nm$) & $K_w$ ($Hz$) \\
        \hline
        \multirow{4}{*}{Curved surfaces} & PTFE ($\phi = 0.14$) & $135^{\circ}$ & 1.68 & $5.67.10^{5}$ \\
        & PTFE ($\phi = 0.18$) & $131^{\circ}$ & 1.30 & $2.80.10^{6}$ \\
        & PTFE ($\phi = 0.23$) & $123^{\circ}$ & 1.02 & $3.35.10^{6}$ \\
        & Smooth PTFE & $103^{\circ}$ & 0.39 & $4.80.10^{7}$ \\
        \hline

        \multirow{3}{*}{Flat surfaces} & PTFE ($\phi = 0.12$) & $139^{\circ}$ & 2.58 & $4.47.10^{5}$ \\
        & PTFE ($\phi = 0.14$) & $135^{\circ}$ & 1.47 & $2.75.10^{6}$ \\
         & PTFE ($\phi = 0.20$) & $123^{\circ}$ & 1.36 & $1.75.10^{6}$ \\

        \hline
    \end{tabular}
    \caption{The MKT fitting parameters used for dash lines in figure \ref{theta_V_SH}.}\label{tab3}
    \end{table}

    \begin{figure}[h!]
    \includegraphics[scale = 1]{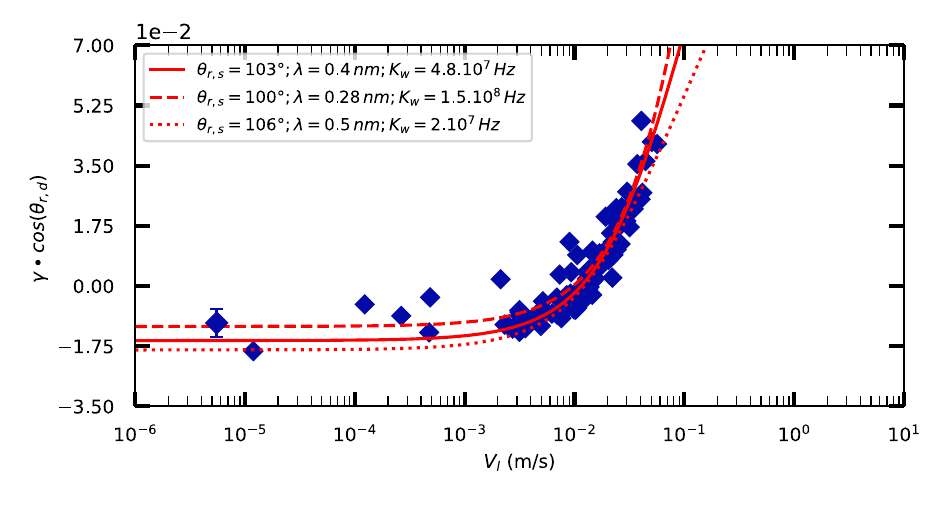}
    \caption{Receding contact angle as a function of contact line velocity adjusted by the molecular model where the value of the static contact angle $\theta_{r,s}$ is fixed at $100 ^{\circ}$, $103 ^{\circ}$ and $106 ^{\circ}$ to show the influence on the parameters $\lambda$ and $K_w$.}
    \label{Fit_variation}
    \end{figure}  
    
\end{document}